\documentclass[aip,a4,onecolumn]{revtex4-2}
\usepackage{amssymb}
\usepackage{amsmath}
\usepackage{color}
\usepackage{float}
\usepackage{graphicx}
\usepackage{hyperref} 
\usepackage{comment}
\usepackage{subcaption}

\begin{document}

\title{Environment-Mediated Long-Ranged Correlations in Many-Body System}

\author{Meng Xu}
\email{meng.xu@uni-ulm.de}

\author{J. T. Stockburger}
\email{juergen.stockburger@uni-ulm.de}

\author{J. Ankerhold}
\email{joachim.ankerhold@uni-ulm.de}
\affiliation{
Institute  for Complex Quantum Systems and IQST, Ulm University - Albert-Einstein-Allee 11, D-89069  Ulm, Germany}

\date{\today}

\begin{abstract}
Quantum states in complex aggregates are unavoidably affected by environmental effects, which typically cannot be accurately modeled by simple Markovian processes. As system sizes scale up, nonperturbative simulation become thus unavoidable but they are extremely challenging due to the intimate interplay of intrinsic many-body interaction and time-retarded feedback from environmental degrees of freedom. In this work, we utilize the recently developed Quantum Dissipation with Minimally Extended State Space (QD-MESS) approach to address reservoir induced long-ranged temporal correlations in finite size Ising-type spin chains. For thermal reservoirs with ohmic and subohmic spectral density we simulate the quantum time evolution from finite to zero temperature. The competition between thermal fluctuations, quantum fluctuations, and anti-/ferromagnetic interactions reveal a rich pattern of dynamical phases including dissipative induced phase transitions and 
spatiotemporal correlations.
\end{abstract}

\maketitle
\newpage

\section{Introduction}
The study of quantum many-body systems often transcends the capabilities of analytical methods \cite{mahan2013many}. Furthermore, traditional numerical simulations confront an intrinsic limitation: the exponential surge in computational resources required as system size increases, rendering them impracticable for large-scale systems. This intricate landscape underpins the potential computational superiority of quantum simulators \cite{georgescu2014quantum} for uncovering the underlying principles of quantum many-body phenomena, especially in areas where classical simulations falter, including quantum magnetism, spin glasses, spin liquids, many-body localization, and quantum phase transitions.

However, the journey towards realizing quantum simulators is fraught with challenges intrinsic to quantum engineering. Notably, quantum many-body systems are susceptible to disturbances from device noise (e.g., charge noise as well as electromagnetic noise) and quantum measurements \cite{preskill2018quantum,forn2019ultrastrong,frisk2019ultrastrong,blais2021circuit}, which becomes more complicated in the scaling of system size. The nature and interaction of this noise with the system can lead to exchange of energy, particles, and informations, significantly impeding the control of the system's coherence and the accurate initial state preparation. Understanding the tolerance to noise of specific quantum simulator run on specific devices is important for determining the feasibility of quantum computing in the current noisy intermediate scale quantum era \cite{preskill2018quantum}. One primary task directly related to fundamental physical properties is understanding the interplay between the environmental effects and the dynamics of interacting many-body systems, as their dynamics exhibits a wide range of features not found in equilibrium or closed systems \cite{diehl2008quantum,verstraete2009quantum,diehl2010dynamical,garrahan2010thermodynamics,weimer2010rydberg,barreiro2011open,ates2012dynamical,cai2013algebraic,preskill2018quantum,krinner2018spontaneous,altman2021quantum,weimer2021simulation,harrington2022engineered}. From a more practical perspective, this interplay is essential in the realm of quantum engineering, where accurate predictions are the pre-requisite for further progress in fields such as quantum information processing \cite{majer2007coupling,ma2019dissipatively,ahn2023modification,sivak2023real,mi2024stable}. 

Despite their rich physics, open quantum many-body systems and in particular interactin spin chains remain vastly understudied, mostly due to the challenges in modeling non-Markovian processes consistently also at low temperatures. Investigating a finite-sized quantum many-body system in contact with a non-Markovian thermal reservoir, one typically encounters a dense and complex Hamiltonian spectrum that precludes the separation of the system's degrees of freedom into distinct energy or time scales, thereby leading to the breakdown of the Born-Markov approximation. This complexity poses a significant challenge in formulating the Lindblad generator that can thermodynamically consistent in predicting nonequilibrium steady-state\cite{gonzalez2017testing,jaschke2019thermalization}. Consequently, the intricate dynamics of open quantum many-body systems often demand a nonperturbative description, for accurately capturing the time-retarded impact of thermal reservoirs onto many-body quantum systems \cite{werner2005phase,schehr2006strong,nakamura2021open,dani2021quantum,makri2021small,de2021quantum,flannigan2022many,bose2022multisite,weber2022dissipation,fux2023tensor,wang2023real,ke2023tree}. 

In this paper, our research focus on an open spin system comprising the transverse field Ising model, subjected to {\em ohmic or subohmic noise}, which displays long-term correlations with algebraic decay. A key aim is to uncover the interplay between the effects of low-frequency noise, bath induced entanglement, coherence, and many-body systems. Providing an answer to this questions would help to understand, e.g., how dynamical quantum phase transition \cite{weiss12} induced in quantum many-body systems by the coupling to noise. In further step, this investigations would enhance understanding of the impacts of $1/f$-noise \cite{dutta1981low,balandin2013low,david2023evolution,paladino2014noise,quintana2017observation} within quantum engineering. Additionally, this investigation encompasses the implementation and optimization of manageable dissipation, a key component also pivotal to the progress in quantum engineering, e.g., generating large-scale multiqubit entaglement \cite{song2017qubit, song2019generation,seetharam2022correlation,parmee2018phase,lledo2019driven,schmi13}. However, the physical challenges in the model setting arise primarily from the scale-free noise spectra at low frequencies, arising from either zero-point quantum fluctuations or thermal fluctuations. The fluctuation-dissipation theorem dictates that these come with equally low-frequence, i.e., sluggish, dynamical response of the environment~\cite{weiss12}. Consequently, there are synergistic effects in the non-Markovian dynamics between thermal fluctuations, quantum fluctuations, and ferromagnetic interactions in quantum many-body systems. These are at best rudimentarily understood at present.

We use the paradigm of Quantum Dissipation with Minimally Extended State Space (QD-MESS) \cite{xu2023universal,vadimov2023nonlinear} in order to address these challenges, and specifically to explore interpaly between thermal fluctuations, quantum fluctuations, and ferromagnetic interactions in quantum many-body systems. This propagation method augments the dynamical state just enough to allow time-local deterministic equations of motion, thereby optimizing computational efficiency while capturing the essential physics accurately for all system observables and correlation functions. In contrast to conventional master equations, its non-unitary terms depend only on the characteristics of the reservoir and the system-reservoir coupling, but not on the level structure of the system proper. Our simulations offer invaluable insights into the control of localization-delocalization dynamics, and potentially the preparation of quantum many-body initial state, e.g., multipartite entangled states.  

The organization of this paper is as follows: In Section II, we model the open quantum system and present the QD-MESS approach. Detailed information about the open quantum many-body system, including qualitative analysis and numerical simulation results, is provided in Section III. We summarize the research findings in Section IV.

\section{Modeling open quantum system}
The open spin systems can be modeled as system plus bath \cite{weiss12}, where the total Hamiltonian ($\hbar = k_B = 1$) reads
\begin{equation}\label{Eq:htot}
    \hat{H} = \hat H_\mathrm s + \hat H_\mathrm {sb} + \hat H_\mathrm b \;\; .
\end{equation}
Here the system of interest $\hat H_\mathrm s$, is embedded in a heat bath with bulk properties, i.e.,  many degrees of freedom, characterized by Hamiltonian $\hat{H}_{\rm b}$, governing free fluctuations. We assume a Gaussian environment, i.e., an environment with fluctuations where cumulants of higher order than two can be neglected.

A generic Hamiltonian representation of such an environment consists of a large, possibly infinite number of harmonic oscillators with Hamiltonian
\begin{equation}\label{Eq:hbath}
    \hat{H}_{\rm b} = \sum_{j} \left( \frac{\hat{p}_j^2}{2m_j} + \frac{1}{2} m_j \omega_j^2 \hat{x}_j^2 \right) \;\;,
\end{equation}
where \(p_j\), \(m_j\), and \(x_j\) represent the momentum, mass, and position of the \(j\)-th oscillator, respectively, and \(\omega_j\) the natural frequency. Alternatively, the environment can be conceived as modes of a quantum field; then $x_j$ and $p_j$ should be considered quadratures of harmonic modes labelled by $j$.

The interaction between the many-body system and the reservoir excitations can generally vary in a non-trivial way from mode to mode, in particular, if both system and environment have large spatial extension. Here we concentrate on the case where the system is either small compared to any relevant wavelengths of the environment or more generally an environment where every reservoir mode interacts with different locations of the system equally and without site-dependent phase delays.

In these cases one may assume an linear, separable form for $\hat{H}_{\rm sb}$ with 
\begin{equation}\label{Eq:hsb}
    \hat{H}_{\rm sb} = -\hat{q}_{\rm s} \sum_j c_j\hat{x}_j  = -\hat{q}_{\rm s}\hat{X}_b \;\;,
\end{equation}
where \(c_j\) are the coupling constants, and \(\hat{q}_{\rm s}\) in our present research denotes a collective operator (see below) that acts on the state space of the many-body system. This type of collective interaction between spins and an environment \cite{vadim21}is also reminiscent of Dicke physics~\cite{garra11}. 

It should be noted that the assumption of a collective coupling to a common bath in our study aims to study the interplay between dissipation and bath-induced long-range interactions according to recent experimental developments. Various setups have been designed to realize spins coupled to a common bath, such as cold-atom systems \cite{recati2005atomic,gadway2010superfluidity}, atom-cavity systems \cite{mlynek2014observation, pandya2021microcavity, plumhof2014room}, and artificial qubits linked to a bus resonator \cite{song2017qubit, song2019generation}. In these contexts, a common bath model offers a realistic approximation.

Now, since thermal fluctuations follow the fluctuation-dissipation theorem, the reservoir's fluctuation spectrum 
\begin{equation}
    S_\beta(\omega) = \int_{-\infty}^{+\infty} dt\, C(t)\, {\rm e}^{i\omega t} 
    =  \int_{0}^{+\infty} dt\, C(t)\, {\rm e}^{i\omega t} + c.c.\;\;
\end{equation}
can be obtained from the inverse reservoir temperature $\beta=1/T$ and a spectral density characterising the dynamical response of the reservoir, i.e., $S_\beta(\omega) = n_\beta(\omega) J(\omega)$. Here, the spectral density 
\begin{equation}
J(\omega) = \pi\sum_{j=1}^{+\infty} \frac{c_j^2}{m_j\omega_j} \delta(\omega - \omega_j)
\end{equation}
is introduced as an anti-symmetric function and $n_{\beta}(\omega) = 1/[1-\exp(-\beta\omega)]$ is the Bose distribution. In the continuum limit, $J(\omega)$ is typically a smooth function; frequently a power law with an exponential cutoff is assumed,
\begin{equation}
\label{eq:Jpow}
J(\omega) =  \pi\alpha\omega_c^{1-s}\omega^s \mathrm{e}^{-\omega/\omega_c}
\end{equation}
for $\omega > 0$. In parallel, the time domain bath correlation function reads: 
\begin{equation}\label{Eq:fluctuations_coth}
\begin{split}
C(t) = \langle \hat{X}_b(t)\hat{X}_b \rangle_\beta 
&= \frac{1}{2\pi}\int_{-\infty}^{+\infty}d\omega\, S_{\beta}(\omega)\, \mathrm{e}^{-i\omega t} \\
&= \frac{1}{2\pi}\int_0^{+\infty} d\omega\, J(\omega)
\left( \coth\frac{\beta\omega}{2} \cos(\omega t) - i \sin(\omega t) \right)\,\, .
\end{split}
\end{equation}

\section{Quantum dissipation with minimally extended state space (QD-MESS)}\label{sec:qdmess}
A very convenient framework to formulate Gaussian quantum dissipation {\em non-perturbatively} is the path integral representation as pioneered by Feynman and Vernon \cite{feynman63} which since then has been extensively used \cite{weiss12}. The reduced density operator $\hat{\rho}_\mathrm s(t)$ is expressed as a functional integral over paths supported by a Keldysh contour,
\begin{equation}\label{Eq:path-integral}
  \rho_\mathrm s^\pm(t)=\int\mathcal{D}[q_\mathrm s^+, q_\mathrm s^-] \, \mathcal{A}_\mathrm s[q_\mathrm s^+,q_\mathrm s^-] \, \mathcal{F}[q_\mathrm s^+,q_\mathrm s^-]\, \rho_\mathrm s^\pm(0) \;\;.
\end{equation}
Here,  the bare action factor $\mathcal{A}_\mathrm s[q_\mathrm s^+,q_\mathrm s^-] = \exp\{iS[q_\mathrm s^{+}]-iS[q_\mathrm s^{-}]\}$ captures the quantum dynamics in the absence of a bath with $S[q_\mathrm s^+]$,  $S[q_\mathrm s^{-}]$ the corresponding actions associated with forward and backward paths, $q_{\rm s}^{\pm}(\tau)$, respectively. The influence functional
\begin{equation}\label{Eq:IFsuperop}
\mathcal{F}[q_\mathrm{s}^+, q_\mathrm{s}^-] = \exp \left\{ -\int_{0}^{t} ds \int_{0}^{s} d\tau \, \left[q_\mathrm{s}^{+}(s) - q_\mathrm{s}^{-}(s)\right] \left[C(s-\tau)q_\mathrm{s}^{+}(\tau) - C^{\ast}(s-\tau)q_\mathrm{s}^{-}(\tau) \right] \right\} \,\,
\end{equation}
parameterized by system state encodes bath effect on the local system. The endpoints of these paths appear explicitly when matrix elements (conventionally with respect to position) of the initial and final densities are considered. Here and in the sequel, we use a shorthand notation and indicate this dependence by superscripts $\pm$ of the respective densities $\langle \rho^{\pm}\rangle = \langle q^- |\rho|q^+\rangle$.

It can be seen that the only information required about thermal environments are the correlation function $C(t)$. Knowledge about actual microscopic degrees of freedom is not necessary which implies wide applicability and flexible technique treatment. In physical settings where dissipative effects lead to a strong suppression of quantum coherence, a direct numerical evaluation of Eq. (\ref{Eq:path-integral}) through path-integral Monte Carlo methods can be efficient~\cite{muhlb04,muhlbacher2005nonequilibrium,kast13}. When quantum coherence is present, however, these Monte Carlo methods suffer from a sign problem, which can make them prohibitively expensive.

A notable feature of influence functionals is their ability to describe arbitrary long-ranged self-interactions in time of the system paths. This feature poses significant challenges in deriving an exact generalized time-local master equation. Time locality can be achieved when an extended state is propagated which consists of the reduced density matrix and additional degrees of freedom. For the contruction of this extended state one may leverage the approach that the bath correlation function can be very accurately approximed by a summation of exponentials \cite{xu2022taming,chen2022universal},
\begin{equation}
    C(t\geq 0) = \sum_{k=1}^K d_k\, \mathrm{e}^{-z_k t} \qquad \mathrm{Re}\{z_k\} > 0 \;\;,
\end{equation}
the influence functional (\ref{Eq:IFsuperop}) can be unravelled into {\em coherent state path integral} \cite{xu2023universal}, i.e.,
\begin{align}\label{Eq:fullaction}
       \mathcal{F}[q_\mathrm{s}^+,q_\mathrm{s}^-] = \prod_k\int\mathcal{D}[\phi_k^\ast,\phi_k;\psi_k^\ast,\psi_k]\, \exp\left\{ iS_k[\phi_k^\ast,\phi_k;q_{\rm s}^+, q_{\rm s}^-]\right\} \exp\left\{ i\bar{S}_k[\psi_k^\ast,\psi_k;q_{\rm s}^+, q_{\rm s}^-]\right\} \;\;
\end{align}
with vacuum state boundary conditions \cite{xu2023universal}, i.e.,
\begin{equation}
\label{eq:initialcond}
    \phi_k(0) = \phi_k(t) = \psi_k(0) = \psi_k(t) = 0 \;\;,
\end{equation}
and actions
\begin{subequations}
\begin{equation}
    S_k[\phi_k^\ast,\phi_k;q_{\rm s}^+, q_{\rm s}^-] = \int_0^td\tau \left[\phi_k^\ast i\partial_\tau \phi_k -iH_k(\phi_k^\ast,\phi_k,q_{\rm s}^+, q_{\rm s}^-) \right]
\end{equation}    
\begin{equation}
    \bar{S}_k[\psi_k^\ast,\psi_k;q_{\rm s}^+, q_{\rm s}^-] = \int_0^td\tau \left[\psi_k^\ast i\partial_\tau \psi_k -i\bar{H}_k(\psi_k^\ast,\psi_k,q_{\rm s}^+, q_{\rm s}^-) \right]
\end{equation}
\end{subequations}
with non-Hermitian Hamiltonians
\begin{subequations}\label{Eq:CS}
\begin{equation}
    H_k(\phi_k^\ast,\phi_k) = -z_k \phi_k^*\phi_k - \sqrt{d_k} (q_\mathrm s^+ - q_\mathrm s^-)\phi_k + \sqrt{d_k} q_\mathrm s^+\phi_k^*  \;\;\;
\end{equation}
\begin{equation}
    \bar{H}_k(\psi_k^\ast,\psi_k) = -z_k^* \psi_k^*\psi_k - \sqrt{d_k^\ast}(q_\mathrm s^- - q_\mathrm s^+)\psi_k + \sqrt{d_k^\ast} q_\mathrm s^-\psi_k^* \;\;.
\end{equation}
\end{subequations}

\begin{figure}[!ht]
    \centering
    \includegraphics[width=\textwidth]{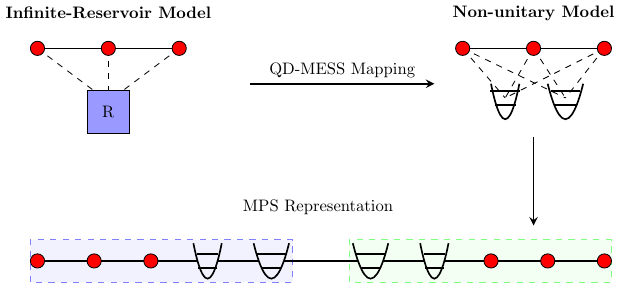}
    \caption{
        The pathway from the infinite reservoir model Hamiltonian (Eq. \ref{Eq:htot}) to an effective non-unitary model where the dynamics are governed by the QD-MESS equation (Eq. \ref{Eq:standard-heom-liouville-fock}). The state of the total degrees of freedom in the QD-MESS is approximated with a matrix product state (MPS). We align the dual space degrees of freedom of the many-body system and the \(\{\hat{a}_k^\dagger, \hat{a}_k\}\) degrees in the left (weak blue part), and the tilde ones along with the \(\{\hat{b}_k^\dagger, \hat{b}_k\}\) degrees in the right (weak green part) \cite{xu2023universal}. Here, the open boundary many-body system (red circles) is illustrated with \(L = 3\), and \text{R} denotes the reservoir (blue rectangle). In the QD-MESS, the reservoir is modeled with a minimal number of harmonic oscillators (\(K = 2\) as an illustration) characterized by complex parameters \cite{xu2022taming, xu2023universal}.
    }
    \label{fig:pathway}
\end{figure}

With the introduction of additional path variables in Eq. (\ref{Eq:fullaction}), all action terms become time local when the unraveling (\ref{Eq:fullaction}) is inserted into the full path integral (\ref{Eq:path-integral}). This corresponds to an extended dynamics which can be described conventionally, using linear operators as generators and states taken from a linear space. What is unconventional here is the appearance of action terms involving external sources $(q_{\rm s}^+,q_{\rm s}^-)$, describing a mixed state, coupling to the coherent-state complex paths that describing pure states. Thus, the resulting state space in the Schr\"odinger picture of the dynamics described by Eq. (\ref{Eq:fullaction}) is neither a Liouville space (mixed states) nor a quantum mechanical Hilbert space (pure states). Since the path integral (\ref{Eq:fullaction}) has a forward-backward path structure for the system paths, but not for the coherent-state path variables, the corresponding quantum states form a product space where one factor is the quantum Liouville space of the system, the other the Fock space of a $2K$-mode harmonic system,
\begin{equation}
        \mathsf{\Gamma} = \mathbb{L}_\mathrm s \otimes \mathbb{F}_{2K}\;\,.
\end{equation}
Starting from a system-environment model with an infinite number of environmental modes, one arrives at an equivalent, somewhat more abstract model with a finite, numerically manageable number of non-hermitian bosonic modes, as illustrated in Fig.~\ref{fig:pathway}.

Assigning raising and lowering operators $\hat{a}_k^\dagger$, $\hat{a}_k$, $\hat{b}_k^\dagger$ and $\hat{b}_k$ to the pure-state modes described by the coherent-state paths $\phi_k$, $\phi_k^\ast$, $\psi_k$ and $\psi_k^\ast$, the dynamics of an extended state $\hat{\rho}(t) \in \mathsf{\Gamma}$ in mixed Liouville-Fock state space reads\cite{xu2023universal}
\begin{equation}\label{Eq:standard-heom-liouville-fock}
\begin{split}
    \partial_t \hat{\rho}(t) =& -i[\hat{H}_{\rm s}, \hat{\rho}(t)] - \sum_k \left[z_k\hat{a}_k^\dagger\hat{a}_k + z_k^\ast\hat{b}_k^\dagger\hat{b}_k \right] \hat{\rho}(t) \\
    &+\sum_{k} \left\{\sqrt{d_k}\,\hat{q}_{\rm s}\,\hat{a}_k^\dagger\,\hat{\rho}(t) - \sqrt{d_k}\,[\hat{q}_{\rm s},\, \hat{a}_k\,\hat{\rho}(t)] \right\} \\
    &+\sum_{k} \left\{\sqrt{d_k^\ast}\,\hat{b}_k^\dagger\,\hat{\rho}(t)\,\hat{q}_{\rm s} + \sqrt{d_k^\ast}\,[\hat{q}_{\rm s},\, \hat{b}_k\,\hat{\rho}(t)] \right\}  \;\;.
\end{split}
\end{equation}

Having established the boundary conditions of the coherent-state path integral in (\ref{eq:initialcond}), we conclude that the factorizing initial condition corresponds to an initial condition with all auxiliary bosons in the vacuum state, and tracing out the real reservoir modes corresponds to formally projecting all auxiliary bosonic modes onto their (formal, not physical) vacuum state \cite{xu2023universal}. This projection is analogous to the distinction between physical and auxiliary density matrices in the hierarchical equations of motion method~\cite{tanimura89,xu2022taming,xu2023universal}.

\section{Finite-size Transverse field Ising model}
For the isolated quantum many-body system we consider a one-dimensional (1D) lattice of spins comprising $L$ sites in form of a transverse field Ising model with open boundary conditions governed by the Hamiltonian
\begin{equation}\label{Eq:hsys}
\hat{H}_s(J, \Delta) = J\sum_{l=1}^{L-1} \hat{\sigma}_l^z\hat{\sigma}_{l+1}^z + \Delta\sum_{l=1}^L \hat{\sigma}_l^x \;\;.
\end{equation}
Here $J$ denotes the strength of the nearest-neighbor spin-spin interactions, $\Delta$ is the transverse field coupling strength, and $\hat{\sigma}_l^{z,x}$ are the Pauli matrices at the $l$th site. This model as such is integrable \cite{pfeuty1970the,chakrabarti2008quantum,sachdev2011quantum} and exhibits a quantum phase transition at zero temperature: As $L \to \infty$, a critical field strength $\Delta_c = J$ marks the transition from an anti-($J>0$)/ferromagnetic ($J<0$) phase for $\Delta < \Delta_c$ with $\langle \hat{\sigma}^z \rangle \ne 0$ to a paramagnetic phase for $\Delta > \Delta_c$ where $\langle \hat{\sigma}^z \rangle = 0$, indicative of the interplay of spin-spin interactions and quantum fluctuations.

The above Hamiltonian posseses symmetries with respect to collective rotations in spin space around the $z-$axis induced by the unitary operator $K_{\frac{\pi}{2}}= \exp(i \frac{\pi}{2}\sum_l \hat{\sigma}_l^z)$. One easily shows that $K_{\frac{\pi}{2}}^\dagger \, \hat{H}_{\rm s}(J, \Delta)\,  K_{\frac{\pi}{2}} = \hat{H}_{\rm s}(J, -\Delta)$ which immeadiately implies the symmetry 
\begin{align}
  \hat{H}_{\rm s}(-J, \Delta)\equiv - \hat{H}_{\rm s}(J, -\Delta)=- K_{\frac{\pi}{2}}^\dagger \, \hat{H}_{\rm s}(J, \Delta)\,  K_{\frac{\pi}{2}}  \,\, .
  \label{eq:symmetry}
\end{align} 
In the sequel, we will pay particular attention to the dynamics of the total magnetization $M(t)$  and the total coherence $X(t)$ in  finite chains, i.e. 
\begin{align}
\label{Eq:magni}
 M(t) = \frac{1}{L} \sum_{l=1}^L \langle \hat{\sigma}_l^z \rangle  \;\; \\
 X(t) = \frac{1}{L} \sum_{l=1}^L \langle \hat{\sigma}_l^x \rangle \;\;.
\label{Eq:cohe}
\end{align}
According to the above symmetry Eq.~(\ref{eq:symmetry}), one has for these expection values the relations 
\begin{align}
\label{Eq:symag}
    M(t)\big|_{J,\Delta} &= M(t)\big|_{-J,\Delta} \\
    X(t)\big|_{J,\Delta} &= -X(t)\big|_{-J,\Delta} \;\;.
\label{Eq:sycoh}
\end{align}

\begin{figure}
    \centering
    \includegraphics[width=\textwidth]{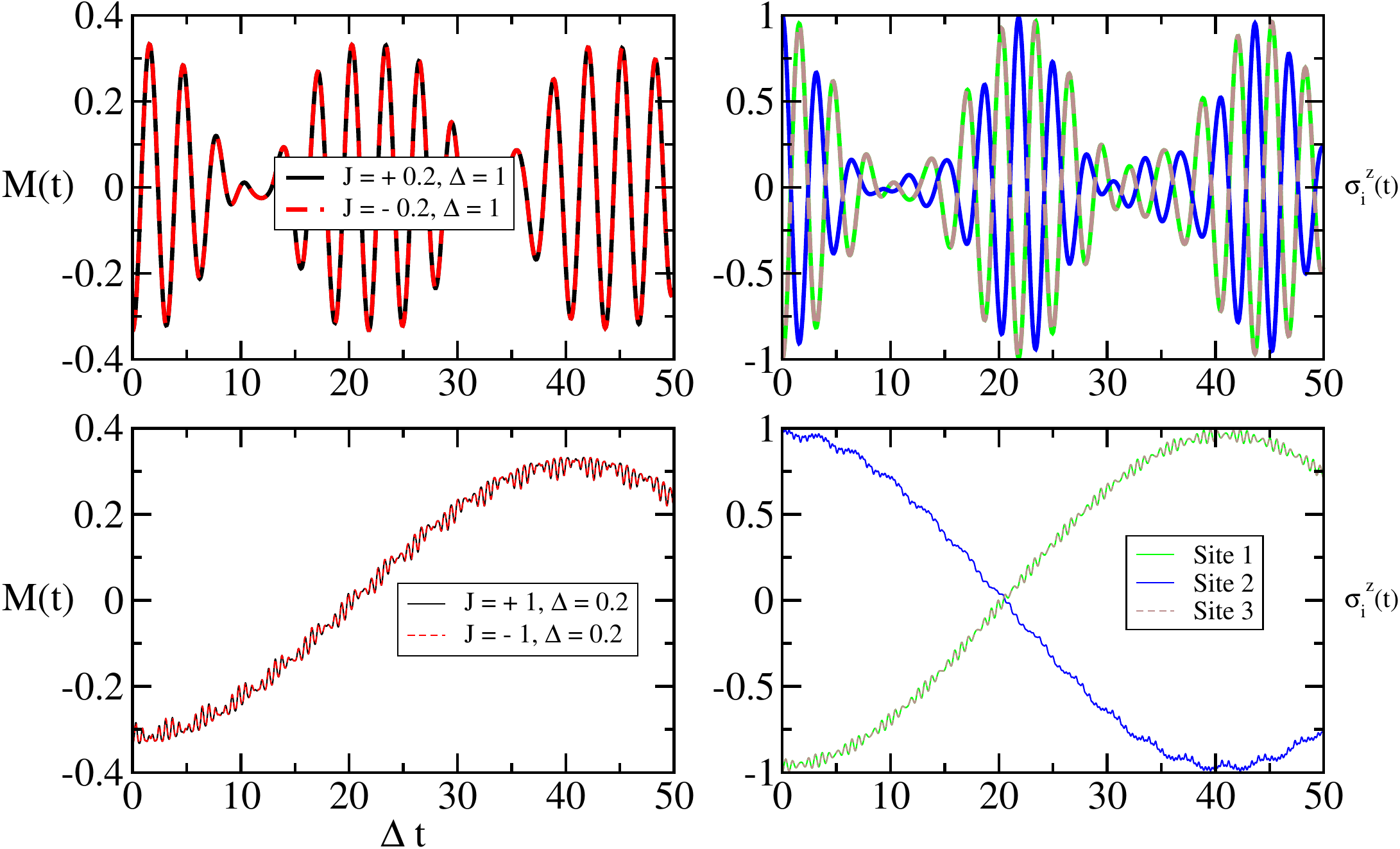}
    \caption{Dynamics of a many-body system in the absence of dissipation. Upper panels: Parameters include spin-spin interaction \(J = \pm 0.2\) and quantum tunneling \(\Delta = 1\). Lower panels: Parameters are \(J = \pm 1\) and quantum tunneling \(\Delta = 0.2\). The left panels display the average magnetization, while the right panels illustrate site-dependent magnetization. The spins ($L = 3$) are initially aligned in the N\'eel state, i.e., $|\downarrow\, \uparrow\, \downarrow\rangle$.}
    \label{fig:sz}
\end{figure}

\begin{figure}
    \centering
    \includegraphics[width=\textwidth]{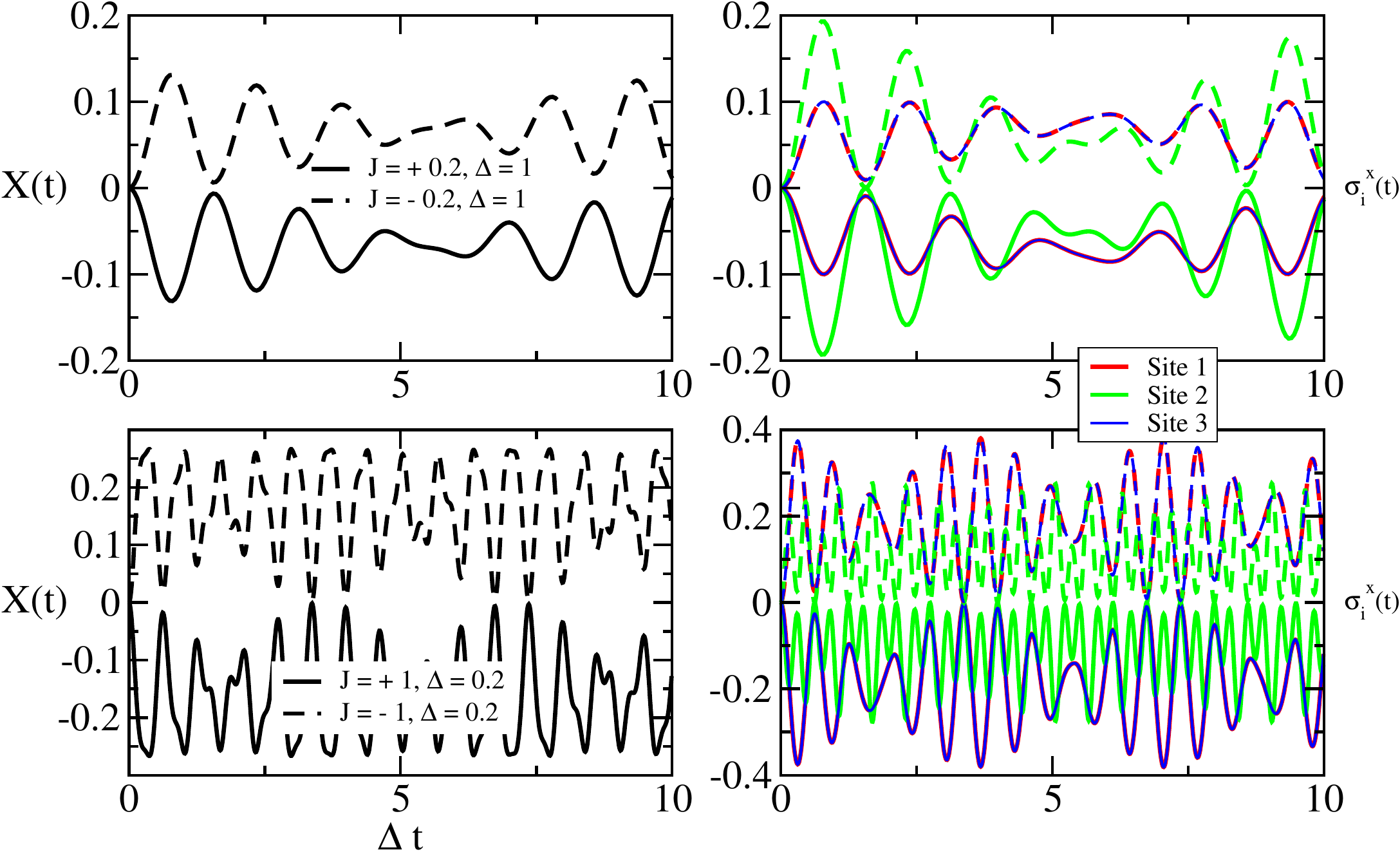}
    \caption{Dynamics of a many-body system, quantified as \(X(t) = \frac{1}{L} \sum_{i=1}^L \langle \hat{\sigma}_i^x(t) \rangle\), in the absence of dissipation. Upper panels: The solid line represents \(J = +0.2\), while the dashed line represents \(J = -0.2\), with quantum tunneling \(\Delta = 1\). Lower panels: The solid line represents \(J = +1\), while the dashed line represents \(J = -1\), with quantum tunneling \(\Delta = 0.2\). The left panels display average transverse magnetization, whereas the right panels show site-dependent transverse magnetization. The spins ($L = 3$) are initially aligned in the N\'eel state, i.e., $|\downarrow\, \uparrow\, \downarrow\rangle$.}
    \label{fig:sx}
\end{figure}

The time evolution of the bare quantum Ising chain is described by the density operator $\hat{\rho}(t)$, following the Liouville equation:
\begin{equation}\label{Eq:sys-Liouville} 
\frac{d}{dt}\hat{\rho}_{\rm s}(t) = -i[\hat{H}_{\rm s}, \hat{\rho}_{\rm s}(t)] \;\;.
\end{equation}
For numerical simulations we employ a tensor product basis of local spin states, denoted by $|q_{\rm s}\rangle = |\sigma_1\rangle\otimes|\sigma_2\rangle\ldots\otimes|\sigma_L\rangle$, as the computing basis, where $\sigma_l \in \{\uparrow,\, \downarrow \}$ and $\hat{\sigma}_l^z|\sigma_l\rangle = \sigma_l|\sigma_l\rangle$. Figures \ref{fig:sz} and \ref{fig:sx} present the spin dynamics for different values of \( J \) and \(\Delta\). It can be seen that the magnetization and coherence in the numerical simulations are consistent with the symmetry analysis as in Eqs. (\ref{Eq:symag}) and (\ref{Eq:sycoh}). 

In fact, the transver Ising model has received particular attention recently in the context of quantum simulators. Experimental implementations of the transverse field Ising model, including artificial spin ice (ASI) array \cite{bingham2021experimental}, trapped atomic ions \cite{kim2011quantum} and therein, neutral atoms \cite{graham2022multi}, and superconducting circuits \cite{kim2023evidence}, etc., act as platforms for quantum simulation, albeit under the assumption of an isolated system. However, the engineering fabrication and operation of quantum devices unavoidably subject these many-body systems to interactions with their environment and to the effects of measurement. This interaction plays multiple roles: it commonly disrupts the coherence of many-body systems, but it also has the positive action in generating multiqubit entanglement which can significantly enhance the efficiency to create large-scale entanglement \cite{song2017qubit,song2019generation,fistul2022quantum}. Consequently, the presence and implementation \cite{mi2024stable} of dissipation critically affects the control and preparation of quantum many-body states. Therefore, it is challenging to qualitatively ascertain whether the effects of dissipation are beneficial or detrimental.

\section{Dissipative Hamiltonian}
The system-reservoir Hamiltonian, as defined in Eq.~(\ref{Eq:htot}) can be recast as ($m_j = 1$):
\begin{equation}\label{Eq:htot2}
    \hat{H} = \hat{H}_{\rm s}(J, \Delta) + \frac{1}{2}\sum_{j}\left\{\hat{p}_j^2 + \omega_j^2 \left(\hat{x}_j - \frac{c_j}{\omega_j^2}\hat{q}_{\mathrm{s}} \right)^2\right\} - \mu \hat{q}_{\mathrm{s}}^2 \,\,,
\end{equation}
where, using Eq.~(\ref{eq:Jpow}), the constant $\mu$ reads
\begin{equation}\label{Eq:hc}
    \mu = \frac{2}{\pi} \int_{0}^{+\infty} d\omega\, \frac{J(\omega)}{\omega} = 2\,\alpha\,\omega_c\,\frac{\Gamma(s + 1)}{s}\;\;
\end{equation}
and where
\begin{equation}\label{Eq:qcollec}
    \hat{q}_s = \sum_l \hat{\sigma}_l^z \;\;.
\end{equation}
The effects of the system-bath interaction can be further analyzed by applying the polaron transformation
\begin{equation}
    \mathcal{S} = \exp\left\{\frac{1}{2} \Omega\, \hat{q}_{\mathrm{s}} \right\}, \quad \Omega = i\sum_{j}\frac{2c_j}{\omega_j^2}\,\hat{p}_j \,,
\end{equation}
so that the total Hamiltonian $\hat{H}$ is mapped onto
\begin{equation}
\label{eq:htile}
    \tilde{H} = \mathcal{S}\hat{H}\mathcal{S}^{-1} = J\sum_{l} \hat{\sigma}_l^z\hat{\sigma}_{l+1}^z + \Delta\sum_l \left[\hat{\sigma}_l^+\mathrm{e}^{i\Omega} +  \hat{\sigma}_l^-\mathrm{e}^{-i\Omega}\right] - \mu\sum_{j,k} \hat{\sigma}_j^z\hat{\sigma}_k^z +  \frac{1}{2}\sum_j \left[\hat{p}_j^2 + \omega_j^2 \hat{x}_j^2 \right] 
\end{equation}
with $\sigma_l^{\pm} = \frac{1}{2}(\sigma_x \pm i\sigma_y)$. 

Apparently, the symmetry (\ref{eq:symmetry}) discussed above for the bare spin model is broken for the dissipative case (\ref{eq:htile}) since  
\begin{align}
\label{Eq:heff_symmetry}
\tilde{H}(-J, \Delta, -\mu)= - K_{\frac{\pi}{2}}^\dagger \, \tilde{H}(J, \Delta, \mu)\,  K_{\frac{\pi}{2}}\,\,
\end{align}
includes the reorganization energy $\mu$. However, as intrinsic quantity $\mu$ is always positive and cannot be chosen freely. 

Further, the excitation of a local site spin results in the simultaneous polarization of the adjacent bath into a coherent state $|\Omega\rangle = e^{i\Omega}|{0}\rangle$, leading to a renormalization of the tunneling element \cite{orth2008dissipative,orth2010dynamics,henriet2016quantum,winter2014quantum}. Deep insight can be obtained by solving the Liouville equation $\partial_t \rho(t) = -i[\tilde{H}, \rho(t)]$ up to second order, i.e., using the non-interacting blip approximation (NIBA) \cite{weiss12}. In this domain and for Ohmic bath spectra wtih $s = 1$, the tunneling frequency is renormalized compared to its bare value $\Delta$ to an effective one \cite{weiss12}:
\begin{equation}\label{Eq:teff}
    \Delta_r = \Delta \left(\frac{\Delta}{\omega_c}\right)^{\frac{\alpha}{1-\alpha}} \,\,.
\end{equation}
This renormalization elucidates the mechanism underlying the dissipative quantum phase transition induced by the bath: With increasing dimensionless coupling $\alpha$ the effective tunneling $\Delta_r$ is supressed while at the same time  Ising-type ferromagnetic interactions are enhanced due to the $\mu$-term mediated by the exchange of bosonic excitations at low wave vectors. Consequently, it polarizes the spins analogously to a ferromagnetic phase.

We note in passing that the form  of the Hamiltonian  (\ref{eq:htile}) is conveniently employed for imaginary-time path integrals to explore equilibrium properties at  finite temperature. After integrating out reservoir degrees of freedom  \cite{grabert1988quantum,weiss12}, one obtains an effective Euclidean action
\begin{equation}\label{Eq:seff}
     S_{\rm eff}^{\rm E} = \sum_l \int_0^\beta d\tau\, \left[\Delta\sigma_l^x(\tau) + J \sigma_l^z(\tau)\sigma_{l+1}^z(\tau) \right] - \frac{1}{2}\sum_{j,k}\int_0^\beta d\tau d\tau'\, L(\tau - \tau') \sigma_j^z(\tau) \sigma_{k}^z(\tau') \;\;
\end{equation}
with an imaginary-time bath correlation function
\begin{equation}
    L(t) = \frac{1}{2\pi}\int_{0}^{+\infty}d\omega\, J(\omega) \left[\coth{\frac{\beta\omega}{2}}\cosh{\omega t} - \sinh{\omega t} \right] \;\;.
\end{equation}
 It can be observed that the bosonic degrees of freedom are no longer present in the expression of the spin dynamics. This simplification comes at the cost of introducing long-ranged  spatiotemporal spin-spin correlations, which potentially generate collective effects, e.g., quantum phase transitions. After integration by parts, the non-local term in the Euclidean action depends only on jumps in the spin path, corresponding to the operators $\sigma^{\pm}$, and correlators related to $e^{\pm i\Omega}$. In dynamical simulations, the QD-MESS maps these long-range retardations into an equivalent local dynamics in an extended state space.

\section{Interacting spin dynamics in presence of dissipation}
Specifically, our focus is on the localization dynamics, which are influenced by various parameters including the many-body system size $L$, system-bath coupling strength $\alpha$, bath cutoff frequency $\omega_c$, bath mode distribution parameter $s$, and the temperature $T$. In the simulations, unless otherwise noted, we set the total system in a factorized initial state with all spins aligned in the up position, denoted as $|\uparrow, \ldots, \uparrow \rangle$, and $\rho_{\rm b} = e^{-\beta \hat{H}_{\rm b}}/Z_b$ with $Z_b = \rm{Tr}\, e^{-\beta \hat{H}_{\rm b}}$.

To efficiently propagate the dynamics as outlined in Eq.~(\ref{Eq:standard-heom-liouville-fock}), we employ the time-dependent variational principle (TDVP) algorithm, utilizing the matrix product state (MPS) representation \cite{haegeman2016unifying,lubich15,shi2018efficient}, as illustrated in Fig.~\ref{fig:pathway}, and the matrix product operators for QD-MESS has been detailed \cite{xu2023universal}. According to the parameter settings, the maximal bond dimension is set at $\chi = 120$, and the maximal local bosonic basis is $N_b = 40$ in the MPS representation. The duration of the simulations (using MATLAB R2022b) can range from a few minutes to several days using 1 Intel Xeon Gold 6252 CPU @ 2.1 GHz core.

\begin{figure}
    \centering
    \includegraphics[width=\textwidth]{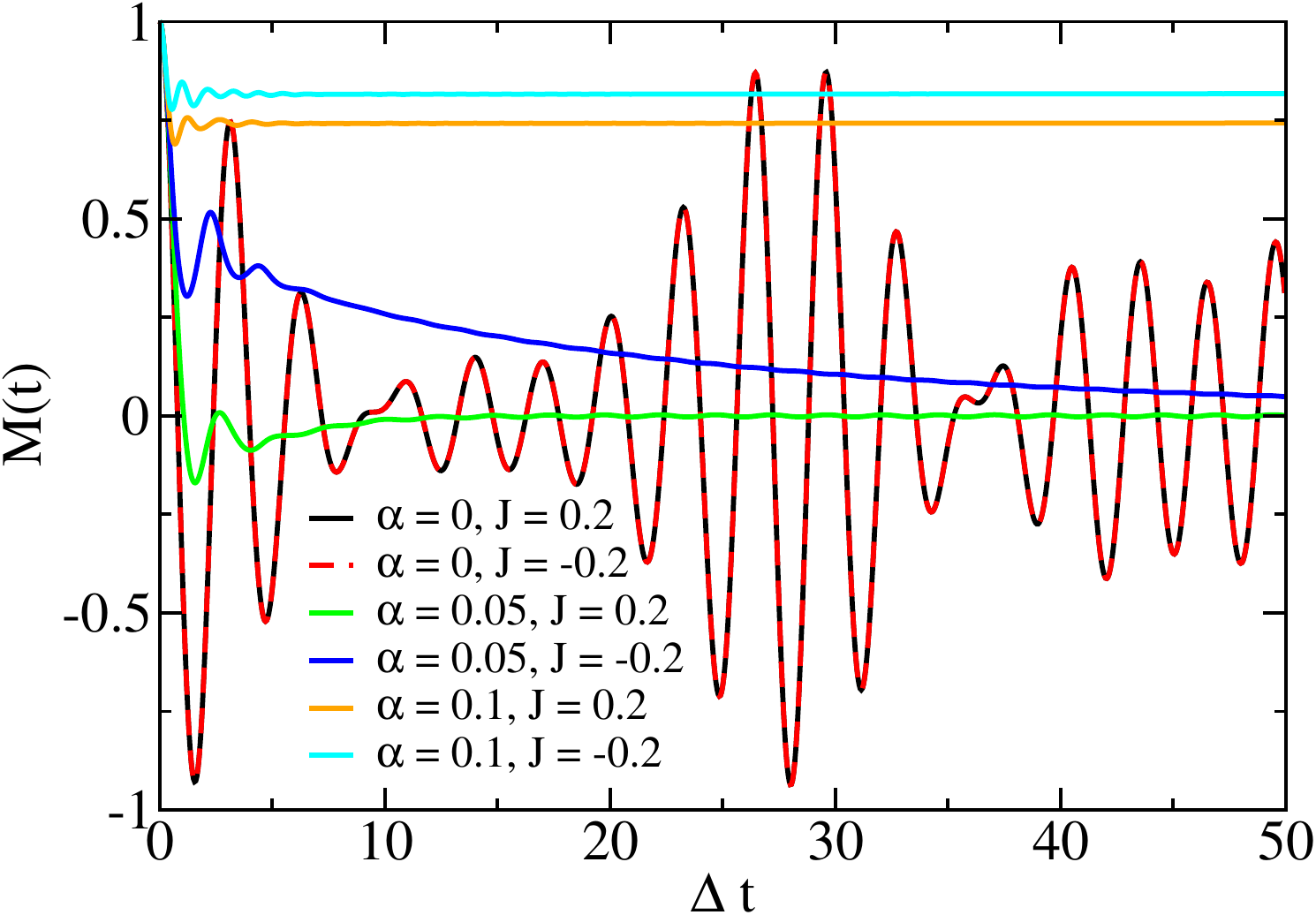}
    \caption{Many-body system dynamics in the presence of non-Markovian noise. Simulation parameters: For the bath, spectral density exponent $s = 1$, cutoff frequency $\omega_c = 10$, and temperature $T = 0$. For the system, lattice size $L = 4$ and transverse field $\Delta = 1$.}
    \label{fig:interplay}
\end{figure}
Figure~\ref{fig:interplay} illustrates the symmetric dynamics for \( J = \pm 0.2 \) which is broken due to the presence of dissipation, i.e., the reorganization energy $\mu$ as in Eq.~(\ref{Eq:heff_symmetry}). As the system-bath coupling becomes stronger, the magnetization dynamics transitions from delocalization into a localization phase where the spins are predominantly frozen in the initial state.

\begin{figure}
    \centering
    \includegraphics[width=\textwidth]{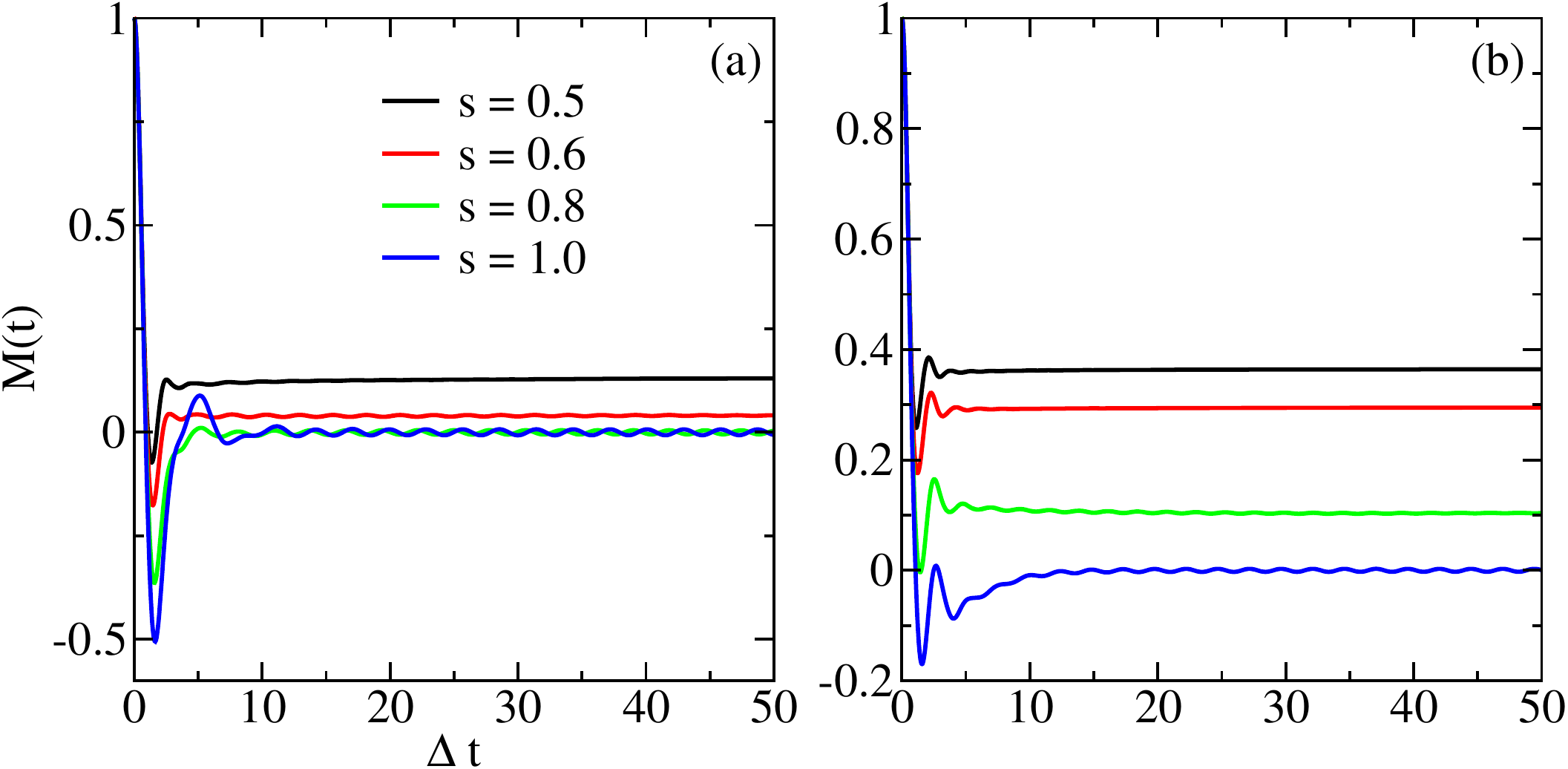}
    \caption{Effects of spectral exponent $s$ onto many-body system dynamics. Simulation parameters are: system-bath coupling $\alpha = 0.05$, bath cutoff frequency $\omega_c = 10$, and temperature $T = 0$. For the system: transverse field $\Delta = -1$, interaction strength $J = 0.2$. Panel (a) depicts a lattice size $L = 3$, while panel (b) shows $L = 4$.}
    \label{fig:leff}
\end{figure}

Figure~\ref{fig:leff} illustrates the influence of the bath parameter $s$ and the system size $L$ on the magnetization dynamics. The subohmic bath serves not only as a medium facilitating interactions among different system sites but also provides a mechanism for long timescale memory effects, significantly impacting the many-body dynamics. The parameter $s$ in the subohmic bath plays a dual role: as $s$ decreases, the influence of the bath's low-frequency modes intensifies, leading to a divergence in the temporal correlations, thus the dynamical quantum phase transition. Additionally, as indicated by Eqs.~(\ref{Eq:hc}) and (\ref{eq:htile}), the ferromagnetic interactions, characterized by $\mu$, become stronger with decreasing $s$. These combined effects lead to the polarization of spins into a localization phase. As the system size $L$ increases, the bath-induced ferromagnetic interaction $\mu$-terms increase rapidly, further enhancing the likelihood of dynamical quantum phase transitions.

\begin{figure}
    \centering
    \includegraphics[width=\textwidth]{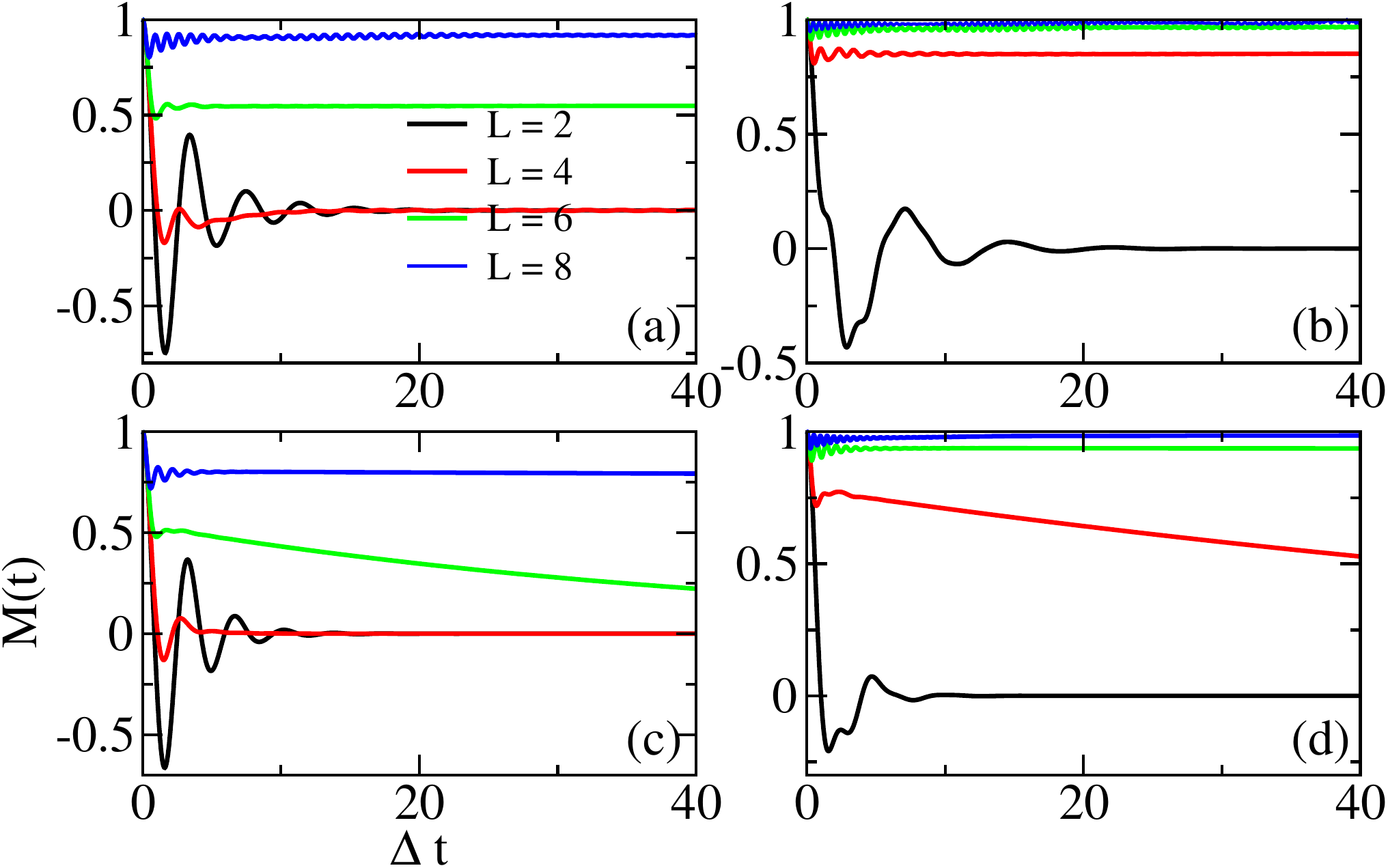}
    \caption{Effects of temperature and bath cutoff frequency onto many-body system dynamics. Simulation parameters are: system-bath coupling $\alpha = 0.05$, spectral density exponent $s = 1$,  transverse field $\Delta = -1$, interaction strength $J = 0.2$. The upper panels correspond to zero temperature, while the lower panels correspond to a finite temperature setting with $\beta = 1$. The left panels illustrate results for a bath cutoff frequency $\omega_c = 10$, whereas the right panels display results for a bath cutoff frequency $\omega_c = 20$.}
    \label{fig:bw}
\end{figure}

Figure~\ref{fig:bw} displays the influence of the cutoff frequency $\omega_c$ and the temperature parameter $\beta$ on the magnetization dynamics. Increases in the cutoff frequency not only renormalize the local tunneling frequency to $\Delta_r \leq \Delta$ (Eq.~(\ref{Eq:teff})), but also enhance the ferromagnetic interactions $\mu$ (Eq.~(\ref{Eq:hc})). Consequently, the spins are frozen in their initial state, an effect that becomes even stronger as the many-body system size increases. However, thermal fluctuations can elevate the tunneling frequency, weakening the localization effects. Therefore, the many-body system dynamics are determined by the interplay between thermal fluctuations, quantum fluctuations, and ferromagnetic interactions.

Of particular interest for dissipative spin chains are the emergence of spatiotemporal long-range correlations induced by the environment. For this purpose, we consider in stationary state the correlator
\begin{equation}
    C_{ij}(t) = \frac{1}{2}\left[\langle \hat{\sigma}_z^i(t) \hat{\sigma}_z^j(0) \rangle + \langle \hat{\sigma}_z^j(0) \hat{\sigma}_z^i(t) \rangle \right]\;\;.
\end{equation}
Note that this necessitates long-time stability of the simulation techniques in order to accurately describe the relaxation dynamics in presence of intricate time-nonlocal spin-reservoir correlations. 
\begin{figure}[h]
    \centering
    \includegraphics[width=\textwidth]{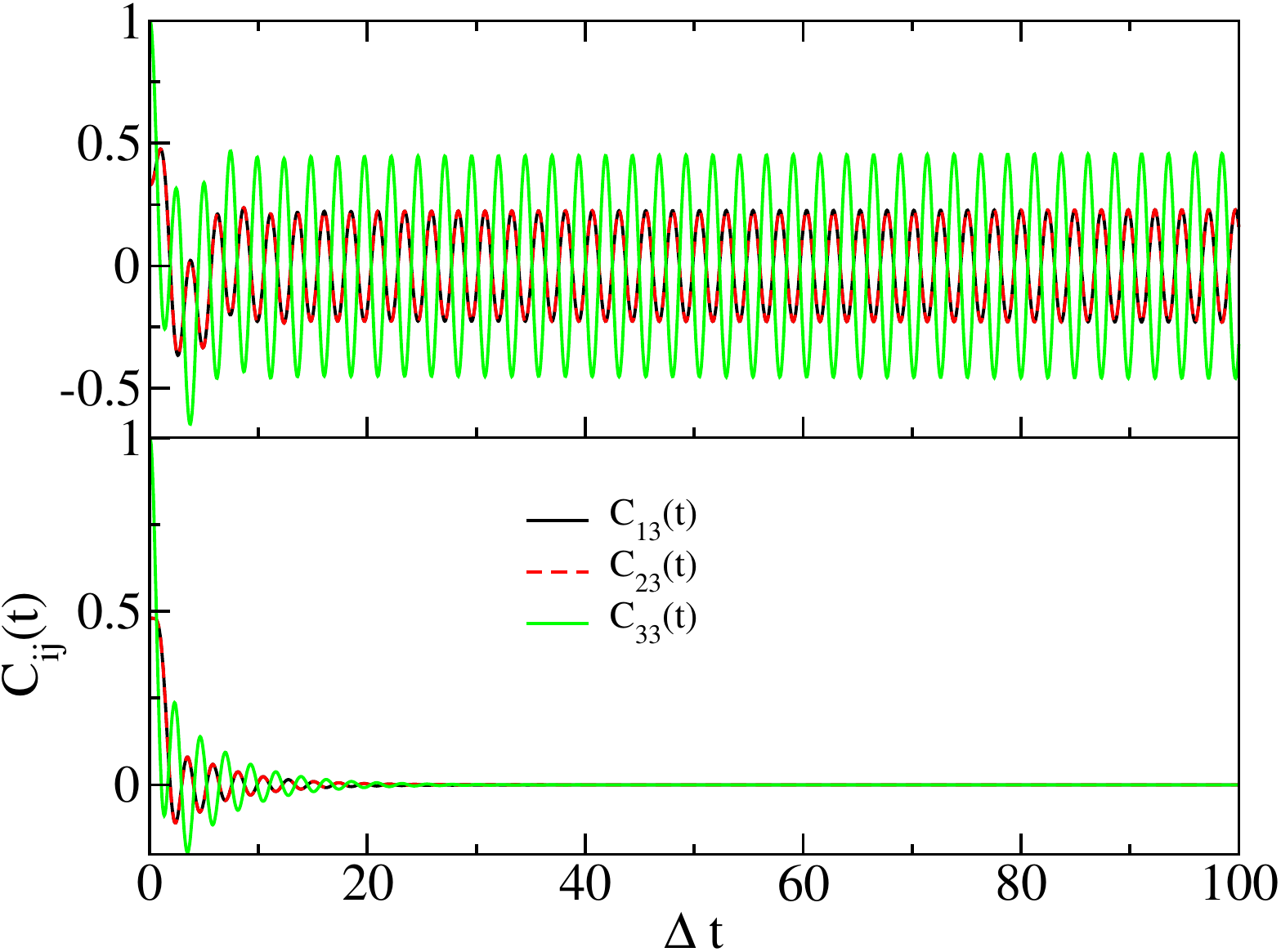}
    \caption{Spatiotemporal correlation functions' dependence on bath temperature. Simulation parameters are: system-bath coupling \(\alpha = 0.05\), spectral density exponent \( s = 1 \), bath cutoff frequency \(\omega_c = 10\), transverse field \(\Delta = 1\), interaction strength \(J = 0\), and many-body system size \( L = 3 \). The upper panels correspond to zero temperature, while the lower panels correspond to a finite temperature setting with \(\beta = 1\).}
    \label{fig:lcorr}
\end{figure}

\begin{figure}[H]
    \centering
    \includegraphics[width=\textwidth]{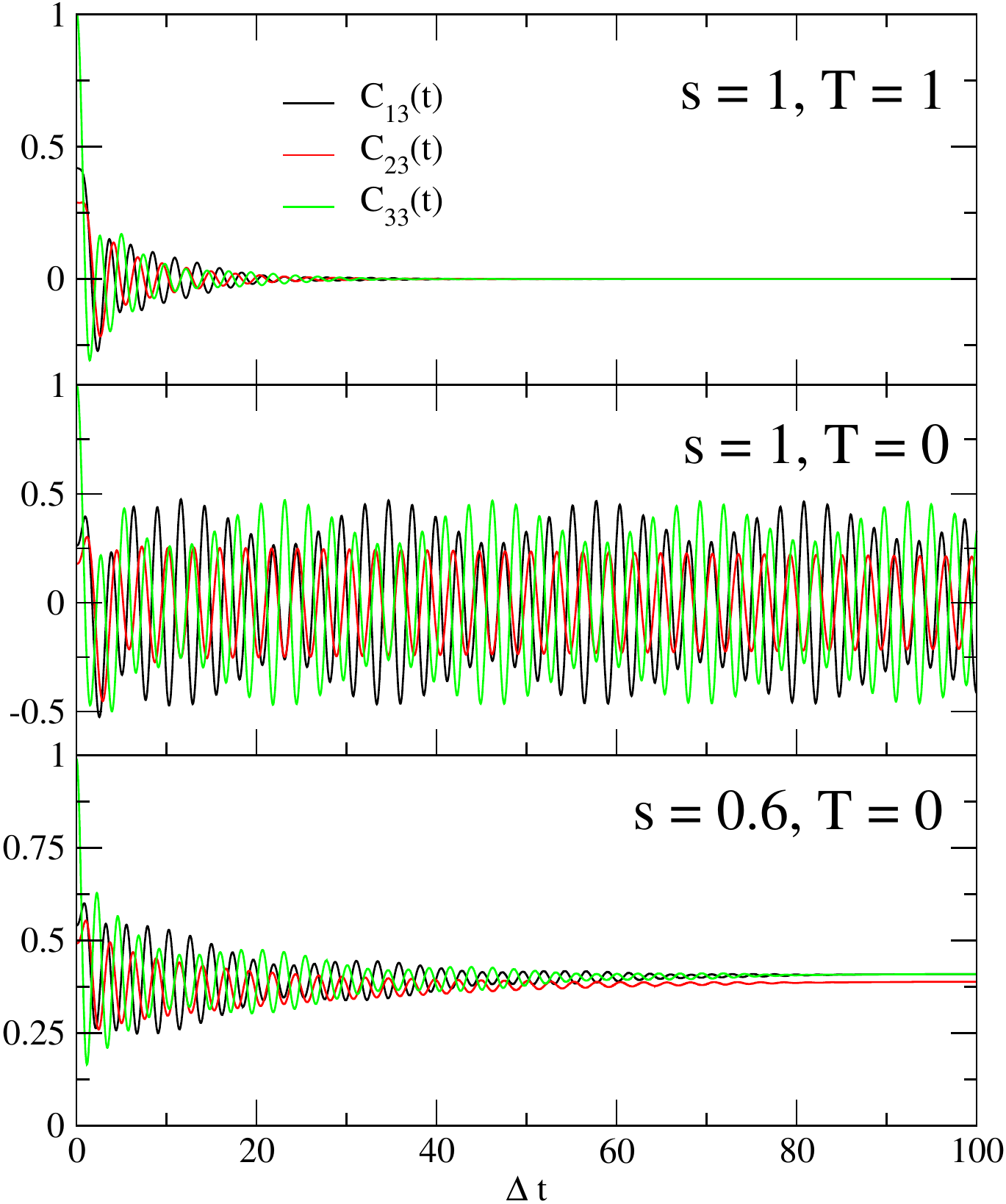}
    \caption{Spatiotemporal correlation functions' dependence on bath temperature and spectral exponent parameter $s$. Simulation parameters are: system-bath coupling \(\alpha = 0.05\), bath cutoff frequency \(\omega_c = 10\), transverse field \(\Delta = 1\), interaction strength $J = 0.2$, and many-body system size \( L = 3 \).}
    \label{fig:corr}
\end{figure}

In Figure~\ref{fig:corr} we show the influence of temperature at  interaction \( J = 0 \). It can be seen that the Ohmic spectral density at zero temperature allows for the mediation of long-ranged correlations and the preservation of quantum coherence for extended periods. However, introducing quantum statistical fluctuations at finite temperatures quickly disrupts these correlations, leading to a rapid decay of the correlation function \( C_{ij}(t) \). This distinction is crucial for understanding and designing quantum many-body systems and their applications, such as resonant or bath mediating long-ranged entanglement.

In the effective Hamiltonian (cf. Eq.~(\ref{eq:htile})), the system's nearest-neighbor interaction characterized by \( J = 0.2 \) seems virtually insignificant compared to the environment-mediated interactions with \( \mu = 1.0 \). This change diminishes the correlation between spin sites 2 and 3, but it introduces long-range interactions between sites 1 and 3 with \( -\mu = -1.0 \). Consequently, the correlation \( C_{13} \) is stronger than \( C_{23} \), as depicted in Fig.~\ref{fig:corr}. Similar to Fig.~\ref{fig:lcorr}, finite temperature thermal fluctuations destroy the system correlations. However, decreasing the exponent parameter \( s \) leads to the domain of dynamical quantum phase transitions, which localize the correlations, maintaining nonzero values.

\section{Summary}
We implement the QD-MESS approach to study the dynamics of an open many-body spin system, specifically targeting at localization dynamics and spatiotemporal correlations where all spins couple to a common ohmic or subohmic reservoir. The bath serves three critical functions: (a) It acts as a medium facilitating interactions among different system sites, exemplified by ferromagnetic interactions; (b) it renormalizes the local tunneling frequency $\Delta$ to $\Delta_r$, which may lead to a dynamical quantum phase transition; (c) it provides thermal fluctuations, which enhance the tunneling frequency and promote many-body system thermalization. The interplay of these effects thus complicates the dynamics of the many-body system. Our studies also demonstrate that system size plays a significant role in the localization-delocalization dynamics. We qualitatively list the dynamical phases dependent on the parameters in Table~\ref{tab:qualitative}.
\begin{table}[ht]
    \centering
        \caption{Qualitative effects of parameter variation on the localization and delocalization in open  quantum many-body systems.}
    \setlength{\tabcolsep}{12pt}
    \begin{tabular}{c|c|c|c|c|c}
    \hline
        Parameters & $L$ & $\alpha$ & $\omega_c$ & $s\in [0,1]$ & $T$ \\
        \hline
         Localization & $\nearrow$ & $\nearrow$ & $\nearrow$ & $\searrow$ & $\searrow$ \\
         \hline
    \end{tabular}
    \label{tab:qualitative}
\end{table}

The potential applications arising from our studies include quantum error-correction and quantum many-body state preparation. Traditional quantum error-correction methods are predicated on error models that rely on Born-Markov assumptions \cite{babu2023quantum}, a limitation that is lifted in the QD-MESS approach. QD-MESS also gives an accurate description of environment-induced self-interactions, a key feature of the results presented here.

Furthermore, in the field of quantum simulations, the preparation of adiabatic states is especially challenging at quantum phase transitions, where many-body energy gaps tend to close \cite{mcclean2018barren}. Additionally, variational quantum algorithms require extensive optimization efforts and often encounter issues known as barren plateaus. In this context, our studies could offer a promising alternative through dissipation engineering to control quantum many-body states.  Lastly, our studies focus on many-body systems that collectively couple to a common bath or cavity modes. This coupling enhances the versatility of operating multiple qubits, which is crucial for practical applications in quantum computing, such as building large-scale quantum networks that could potentially incorporate millions of qubits \cite{majer2007coupling,sillanpaa2007coherent,song2017qubit,xu2018emulating,glaser2023controlled}.

\section*{Acknowledgements}
We acknowledge fruitful discussions with D. Maile,  V. Vadimov, D. Jaschke, and M. M\"ott\"onen. M. X. acknowledges support from the state of Baden-W\"urttemberg through bwHPC (JUSTUS 2). This work has been supported by the German Science Foundation (DFG) under AN336/12-1 (For2724), the State of Baden-W\"urttemberg under KQCBW/SiQuRe, and the BMBF within the project QSolid.

\bibliography{quantum}
\end{document}